\documentclass[11pt,a4paper]{article}

\usepackage{epsfig,amsmath,amssymb,cite}

\begin{document}

\title{Shadow cast by rotating braneworld black holes with a cosmological constant}
\author{Ernesto F. Eiroa$^{1,2,}$\thanks{e-mail: eiroa@iafe.uba.ar}, Carlos M. Sendra$^{1,2,}$\thanks{e-mail: cmsendra@iafe.uba.ar} \\
{\small $^1$ Instituto de Astronom\'{\i}a y F\'{\i}sica del Espacio (IAFE, CONICET-UBA),} \\
{\small Casilla de Correo 67, Sucursal 28, 1428 Buenos Aires, Argentina}\\
{\small $^2$ Departamento de F\'{\i}sica, Facultad de Ciencias Exactas y
Naturales, Universidad de Buenos Aires,} \\
{\small Ciudad Universitaria Pabell\'on I, 1428 Buenos Aires, Argentina} }

\maketitle

\date{}

\begin{abstract}
In this article, we study the shadow produced by rotating black holes having a tidal charge in a Randall-Sundrum braneworld model, with a cosmological constant. We obtain the apparent shape and the corresponding observables for different values of the tidal charge and the rotation parameter, and we analyze the influence of the presence of the cosmological constant. We also discuss the observational prospects for this optical effect.
\end{abstract}


\section{Introduction}

The presence of extra dimensions plays an important role in many gravity theories, within the context of the unification of the physical forces and also in cosmology. Among them are the braneworld cosmological models \cite{branemodels}, motivated by string theory (M-theory) and proposed in order to give an explanation for the so--called hierarchy problem (why the gravity scale is sixteen orders of magnitude greater than the electro-weak scale). The ordinary matter is trapped on a three dimensional space denominated the brane that is embedded in a larger space dubbed the bulk, where only gravity can propagate. The Randall--Sundrum \cite{randall} (second type) model is the simplest of these theories, consisting of a positive tension brane in a bulk with only one extra dimension and a negative cosmological constant. The occurrence of the extra dimensions results in different properties of black holes \cite{kanti}. Primordial black holes formed in the high energy epoch would have a longer lifetime, due to a different evaporation law, and could have a growth of their mass through accretion of surrounding radiation during the high energy phase, increasing their lifetime, so they might have survived up to the present \cite{clanmaj}. High energy collisions in particle accelerators or in cosmic rays could also create black holes \cite{kanti}. In the Randall-Sundrum scenario, the field  equations on the brane \cite{shiromizu} were obtained from five--dimensional gravity, with the help of the Gauss--Codazzi equations. A  spherically symmetric black hole solution on a three dimensional brane was found \cite{dadhich}, characterized by a tidal charge due to gravitational effects coming from the fifth dimension. Non--singular black holes on the brane are also possible \cite{casadio}. A general class of braneworld solutions, describing black holes and wormholes with spherical symmetry was presented \cite{bronnikov}; this work was then extended to the case with non--zero cosmological constant on the brane \cite{molina}. Rotating black hole solutions of the effective field equations on the brane were recently obtained with null \cite{aliev} and also with non--zero cosmological constant \cite{neves}.

The presence of the photon sphere surrounding the horizon of a black hole results in some peculiar features in the behavior of light in its vicinity. The deflection angle diverges at the photon sphere; as a consequence, the light rays coming from a distant source can perform several turns around the black hole before emerging to reach a possible observer. The strong deflection gravitational lensing by black holes was firstly studied in some pioneering works \cite{pioneer} and for the spherically symmetric case, a systematic approach was developed in recent years \cite{sdlsphe}, denominated the strong deflection limit. This method is based on a logarithmic expansion of the deflection angle for light rays passing very close to the photon sphere, from which the positions, the magnifications, and the time delays corresponding to the two infinite sets of the so--called relativistic images can be obtained analytically. The strong deflection limit was used in many cases of interest; among them, non-rotating braneworld black holes were analyzed as gravitational lenses \cite{branelens}. Black hole lensing can be also  studied numerically \cite{numerical}. Kerr black holes were considered as lenses in other articles \cite{kerrlens}; in particular, the strong deflection limit was extended to this case. Another important aspect related to the behaviour of photons in the neighborhood of a black hole is the shadow or apparent shape as seen by a far away observer. The shadows of non-rotating black holes are circles, but rotating ones show a deformation produced by the spin \cite{barchan}. Several researchers have recently studied this topic \cite{shadows1,hioki,shadows2,amarilla,perlick,shadows3,shadowplas}, both in Einstein  theory and in modified gravity. There is strong evidence of the existence of supermassive black holes at the center of most galaxies, including the Milky Way \cite{gillessen17} and the closest one M87 \cite{broderick15}. It is expected that direct observation of black holes will be possible in the near future \cite{observ}, so that the analysis of the shadows will be an useful tool for a better knowledge of astrophysical black holes and also for comparing alternative theories with general relativity.

In this work, we investigate the shadow cast by rotating braneworld black holes with a cosmological constant. The paper is organized as follows: in Sec. \ref{tmetric}, the metric of the black hole with a tidal charge in the Randall Sundrum scenario is introduced, and the equations of motion for photons are obtained. The apparent shape is presented in Sec. \ref{sqt} for different values of the parameters. In Sec. \ref{obs}, the observables are defined and analyzed. Final comments about the results and the future observational possibilities are discussed in Sec. \ref{remarks}. We adopt units such that $G=c=1$, with $G$ the effective four--dimensional gravitational constant and $c$ the speed of light.

\section{Null geodesics on the brane}\label{tmetric}

In the Randall-Sundrum braneworld model, with one brane and a five-dimensional anti--de Sitter bulk with $\mathbb{Z}_2$ symmetry, the induced four-dimensional vacuum field equations on the brane read \cite{shiromizu}
\begin{equation}
G_{\mu\nu}=-\Lambda_{4D}g_{\mu\nu}-E_{\mu\nu},
\end{equation}
where  $G_{\mu\nu}$ is the four-dimensional Einstein tensor associated with metric $g_{\mu\nu}$ of  the brane, $\Lambda_{4D}$ is the effective four--dimensional cosmological constant on the brane, and $E_{\mu\nu}$ is proportional to the projection of the five-dimensional Weyl tensor, which expresses the bulk influence on the brane.  An axially symmetric solution of the field equations on the brane, with a non null effective cosmological constant, was obtained in Ref. \cite{neves}, by assuming a Kerr--Schild--(anti--)de Sitter ansatz.

We start by introducing the metric of the rotating black hole with a cosmological constant found in Ref. \cite{neves}, which in the Boyer--Lindquist coordinates ($t,r,\theta,\phi$) has the form
\begin{eqnarray}
ds^{2} &=& -\frac{1}{\Sigma}\left(\Delta_r-\Delta_{\theta} a^2\sin^2\theta\right)dt^2-\frac{2a}{\Xi\Sigma}\left[\left(r^2+a^2\right)\Delta_{\theta}-\Delta_r\right]\sin^2\theta d\phi dt+\frac{\Sigma}{\Delta_r}dr^2 \nonumber \\
&&+ \frac{\Sigma}{\Delta_\theta}d\theta^2 +\frac{1}{\Xi^2\Sigma}\left[\left(r^2+a^2\right)^2\Delta_{\theta}-\Delta_r a^2\sin^2\theta\right]\sin^2\theta d\phi^2,
\label{metric}
\end{eqnarray}
with
\begin{equation}
\Sigma=r^2+a^2 \cos^2\theta,
\end{equation}
\begin{equation}
\Delta_r=\left(r^2+a^2\right)\left(1-\frac{\Lambda_{4D}}{3}r^2\right)-2Mr+q,
\end{equation}
\begin{equation}
\Delta_{\theta}=1+\frac{\Lambda_{4D}}{3}a^2\cos^2\theta,
\end{equation}
and
\begin{equation}
\Xi=1+\frac{\Lambda_{4D}}{3}a^2,
\end{equation}
where $M$ is the mass, $a$ is the rotation parameter (i.e. the angular momentum per unit mass, $a=J/M$), $q$ is an induced tidal charge on the brane, and $\Lambda_{4D}$ is the four-dimensional brane cosmological constant. This metric is equivalent to the Kerr--Newman--(anti--)de Sitter one in  general relativity, but now the quantity $q$ that replaces the square of the electric charge can be either positive or negative, reflecting the influence of the bulk on the brane. It has been argued that a negative value of $q$ is the physical more natural case \cite{dadhich}. There is a null energy--momentum tensor at the brane and no electromagnetic fields are present there. For this spacetime, the radii of the horizons are determined by the positive solutions of the equation $\Delta_r=0$; for given $M\neq 0$ and $a\neq 0$, the number of them depend on the sign of $\Lambda_{4D}$ and also on the value of $q$. In the case that $\Lambda_{4D}<0$,  when $q<q_1$ the geometry has only an event horizon and  when $q_1<q<q_2$ it has both an internal Cauchy horizon and an external event horizon. The values of $q_1$ and $q_2$, where the number of horizons change, are determined by $M$, $a$ and $\Lambda_{4D}$. In the case that $\Lambda_{4D}>0$, the spacetime has a cosmological horizon with the largest radius, in addition to the event one when $q<q_1$, or to the Cauchy and the event ones when $q_1<q<q_2$. It is worthy to note that, for any $\Lambda_{4D}$, the horizon radius is larger when the tidal charge is negative, so the gravitational effects form the bulk on the brane are amplified in this case. The geometry always presents a ring singularity which is covered by the event horizon when $q$ is not larger than the critical value $q_c=q_2$, corresponding the the extremal case; if this value is exceeded, there is a naked singularity instead of a black hole. For more details, see Ref. \cite{neves}.

In the situation where a black hole is interposed between an observer and an extended background source, not all the photons emitted by the source can reach the observer after being deflected by the black hole gravitational field. The ones with small enough impact parameter, end up falling into the black hole, and give place to a dark region in the sky called the shadow. The contour of the shadow gives the apparent shape of the black hole and it is related to the geodesics of massless particles in the black hole spacetime structure. The geodesics for a given geometry are determined by the Hamilton-Jacobi equation:
\begin{equation}
\frac{\partial S}{\partial\sigma}=-\frac{1}{2}g^{\mu\nu}\frac{\partial S}{\partial x^\mu}\frac{\partial S}{\partial x^\nu},
\end{equation}
where $\sigma$ is an affine parameter along the geodesics and $S$ is the Jacobi action. In the case where $S$ is separable, the Jacobi action can be written in the simple and general form
\begin{equation}
S=\frac{1}{2}m^2\sigma-E t+L\phi+S_r(r)+S_{\theta}(\theta),
\end{equation}
with $m$ the mass of the test particle, $E$ the energy, and $L$ the angular momentum in the direction of the axis of symmetry. The quantities $E$ and $L$ are constants of motion, related to the symmetries of the spacetime and the associated Killing vectors. Considering null geodesics, i.e. $m=0$, and solving the Hamilton-Jacobi equation, the equations of motion for photons propagating in the geometry (\ref{metric}) result
\begin{equation}
\Sigma\frac{dt}{d\sigma}=\frac{r^2+a^2}{\Delta_r}\left[E\left(r^2+a^2\right)-a\Xi L\right]-\frac{a}{\Delta_\theta}\left(aE\sin^2\theta-\Xi L\right),
\end{equation}
\begin{equation}
\Sigma\frac{d\phi}{d\sigma}=\frac{a\Xi}{\Delta_r}\left[E\left(r^2+a^2\right)-a\Xi L\right]-\frac{\Xi}{\Delta_\theta}\left(aE-\Xi L\csc^2\theta\right),
\label{dotphi}
\end{equation}
\begin{equation}
\Sigma\frac{d r}{d\sigma}=\sqrt{R},
\label{dotr}
\end{equation}
and
\begin{equation}
\Sigma\frac{d\theta}{d\sigma}=\sqrt{\Theta},
\end{equation}
with
\begin{equation}
R=\left[(a^2 +r^2)E -a\Xi L\right]^2-\left[\left(L-aE\right)^2+\kappa\right]\Delta_r
\end{equation}
and
\begin{equation}
\Theta=-\left(a E \sin^2\theta-\Xi L\right)^2\csc^2\theta-\left[\left(L-aE\right)^2+\kappa\right]\Delta_{\theta},
\end{equation}
where $\kappa$ is the Carter constant of separation. These equations determine the propagation of light in the spacetime of the rotating black hole on the brane. From the constants of motion $E$, $L$ and $\kappa$, the impact parameters for general orbits around the black hole are defined by the quantities $\xi=L/E$ y $\eta=\kappa/E^2$. The silhouette of the shadow of the black hole is obtained from the orbits of constant $r=r_p$, which satisfy the conditions $R(r)=0=dR(r)/dr$. Solving this system of equations, the impact parameters that determine the contour of the shadow for the braneworld black hole result

\begin{equation}
\xi(r_p)=\frac{\Upsilon(r_p)}{\Gamma(r_p)}
\label{xi}
\end{equation}
and
\begin{equation}
\eta(r_p)=-\frac{\left[\Upsilon(r_p)-2a\Gamma(r_p)\right]\Upsilon(r_p)+a^2\Xi^2\Omega(r_p)}{\Gamma(r_p)^2},
\label{eta}
\end{equation}
where
\begin{equation}
\Gamma(r_p)=a\Xi\left[3M-3r_p+\Lambda_{4D} r_p\left(a^2+2r^{2}_{p}\right)\right],
\end{equation}
\begin{equation}
\Upsilon(r_p)=3\left[M\left(a^2-3r^{2}_{p}\right)+2qr_p+r_p\Xi\left(a^2+r^{2}_{p}\right)\right],
\end{equation}
and
\begin{eqnarray}
\Omega(r_p) &=& 9a^2M^2+3r^{2}_{p}\left[-3a^2\left(1+2\Xi\right)-12\left(q+r^{2}_{p}\right)+4r^{4}_{p}\Lambda_{4D}+\left(a^2+2r^{2}_{p}\right)^2\frac{\Lambda^{2}_{4D}}{3}a^2\right]
\nonumber \\
&& + 18Mr\left[a^2\left(-2+\Xi\right)+2r^2\left(1+\Xi\right)\right].
\end{eqnarray}
For simplicity, all quantities in further calculations are adimensionalized with the black hole mass, i.e. by taking $M=1$.

\section{Black hole shadow}\label{sqt}

In order to obtain the boundary curve of the shadow of the braneworld black hole, we fix the observer position in the domain of outer communication, with Boyer-Lindquist coordinates at $(r_0,\theta_0)$, as it was presented in Ref. \cite{perlick}. At this position, we define the orthonormal tetrad
\begin{equation}
e_0=\left.\frac{\left(r^2+a^2\right)\partial_t+a\Xi \partial_\phi}{\sqrt{\Delta_r\Sigma}}\right|_{(r_0,\theta_0)},
\label{e0}
\end{equation}
\begin{equation}
e_1=\left.\sqrt{\frac{\Delta_\theta}{\Sigma}}\partial_\theta\right|_{(r_0,\theta_0)},
\end{equation}
\begin{equation}
e_2=-\left.\frac{\Xi \partial_\phi+a\sin^2\theta \partial_t}{\sqrt{\Delta_\theta\Sigma}\sin\theta}\right|_{(r_0,\theta_0)},
\end{equation}
and
\begin{equation}
e_3=-\left.\sqrt{\frac{\Delta_r}{\Sigma}}\partial_r\right|_{(r_0,\theta_0)},
\end{equation}
where $e_0$ represents the four velocity of the observer, $e_3$ corresponds to the spatial direction towards the center of the black hole and the combination $e_0\pm e_3$ is tangential to the principal null congruences of the metric. Let $\lambda(s)=(r(s),\theta(s),\phi(s),t(s))$ be the coordinates for each light ray, its tangent vector is given by
\begin{equation}
\dot{\lambda}=\dot{r}\partial_r+\dot{\theta}\partial_\theta+\dot{\phi}\partial_\phi+\dot{t}\partial_t,
\label{lambda1}
\end{equation}
which at the observation event and in terms of the celestial coordinates $\alpha$ and $\beta$ results \cite{perlick}
\begin{equation}
\dot{\lambda}=\gamma\left(-e_0+\sin\alpha\cos\beta e_1+\sin\alpha\sin\beta e_2+\cos\alpha e_3\right).
\label{lambda2}
\end{equation}
\begin{figure}[t!]
\begin{center}
    \includegraphics[scale=0.94]{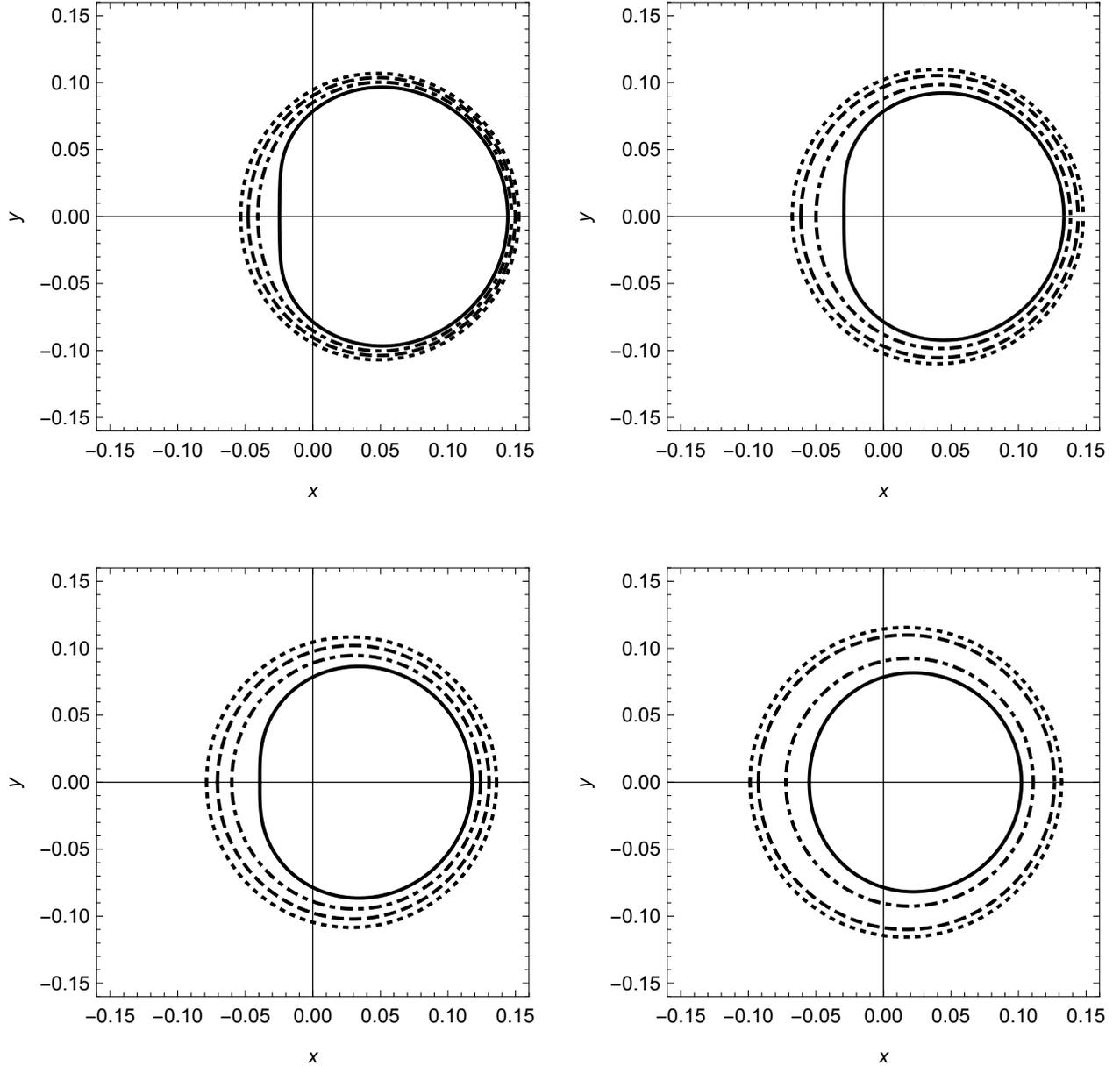}
    \vspace{0.25cm}
    \caption{Apparent shape of the black hole, as seen by an observer at $r_0=50$ and $\theta_0=\pi/2$, with $\Lambda_{4D}=10^{-7}$, for different values of the rotation parameter $a$ and the tidal charge $q$. Upper row, left: $a=0.83$, $q=q_c \approx 0.30$ (full line), $q=0.1$ (dashed-dotted line), $q=-0.1$ (dashed line), and $q=-0.3$ (dotted line). Upper row, right: $a=0.7$, $q=q_c \approx 0.51$ (full line), $q=0.2$ (dashed-dotted line), $q=-0.2$ (dashed line), and $q=-0.5$ (dotted line). Lower row, left: $a=0.5$, $q=q_c \approx 0.75$ (full line), $q=0.4$ (dashed-dotted line), $q=0$ (dashed line), and $q=-0.4$ (dotted line). Lower row, right: $a=0.3$, $q=q_c  \approx 0.91$ (full line), $q=0.5$ (dashed-dotted line), $q=-0.5$ (dashed line), and $q=-0.9$ (dotted line). All quantities are adimensionalized with the mass of the black hole.}
    \label{shadow1}
\end{center}
\end{figure}
\begin{figure}[t!]
\begin{center}
    \includegraphics[scale=0.94]{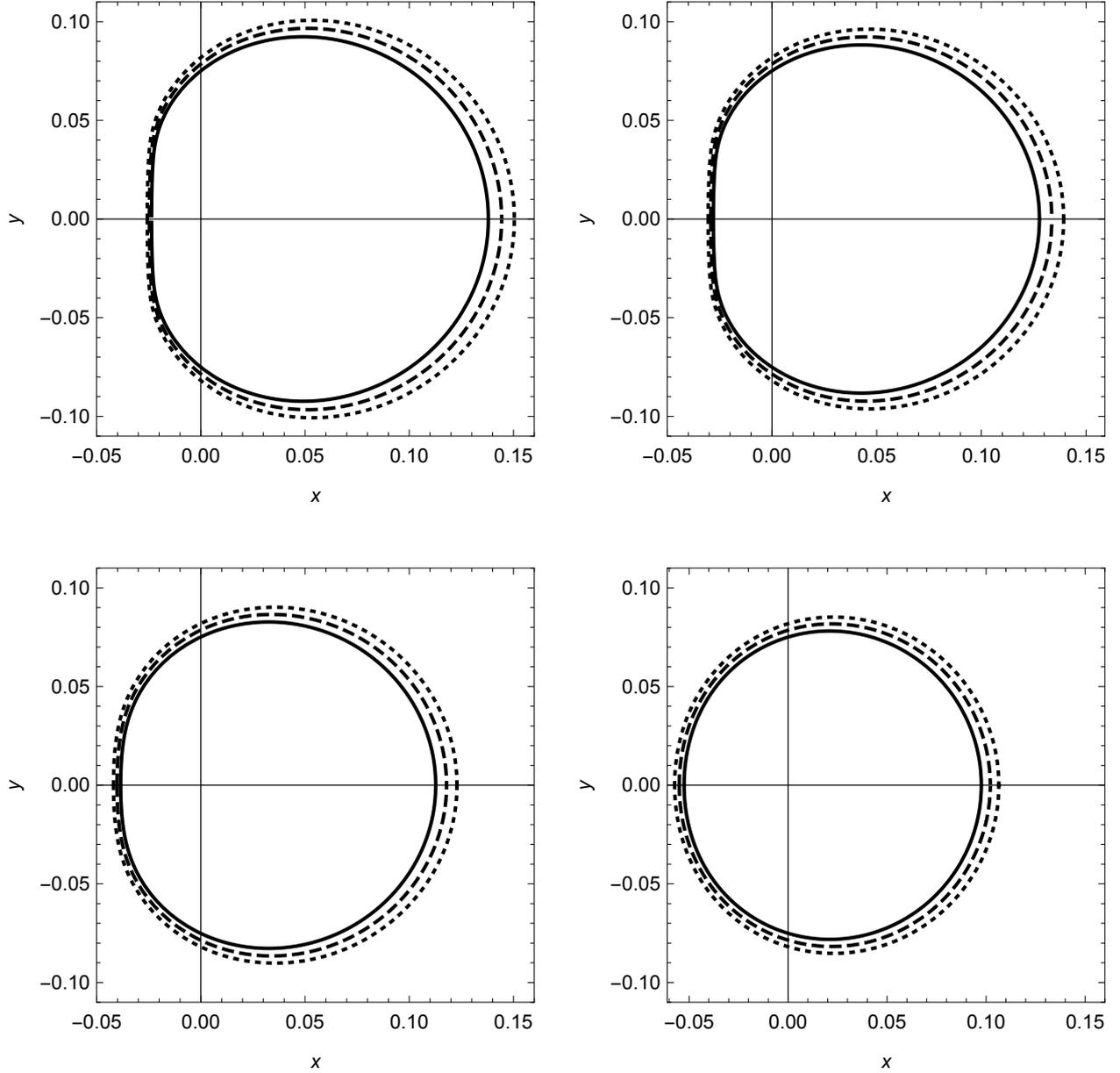}
    \vspace{0.25cm}
    \caption{Shadow of the rotating black hole in the extremal case, for different values of the cosmological constant: $\Lambda_{4D}=-10^{-4}$ (dotted line), $\Lambda_{4D}=10^{-7}$ (dashed line), and $\Lambda_{4D}=10^{-4}$ (full line). The observer is placed at $r_0=50$ and $\theta_0=\pi/2$. Upper row, left: $a=0.83$ and $q_c \approx 0.30$. Upper row, right: $a=0.7$ and $q_c \approx 0.51$. Lower row, left: $a=0.5$ and $q_c \approx 0.75$. Lower row, right: $a=0.3$ and $q_c \approx 0.91$.}
    \label{shadow2}
\end{center}
\end{figure}
The scalar factor $\gamma$ can be easily calculated from Eqs. (\ref{lambda1}) and (\ref{lambda2}):
\begin{equation}
\gamma=g(\dot{\lambda},e_0)=\left.\frac{-E\left(r^2+a^2\right)+a\Xi L}{\sqrt{\Delta_r\Sigma}}\right|_{(r_0,\theta_0)},
\end{equation}
and by comparing the corresponding terms in these expressions, we have that
\begin{equation}
\sin\alpha=\left.\sqrt{1-\left[\frac{\dot{r}\Sigma}{E\left(r^2+a^2\right)-a\Xi L}\right]^2}\right|_{(r_0,\theta_0)}
\end{equation}
and
\begin{equation}
\sin\beta=\left.\frac{\sqrt{\Delta_\theta}\sin\theta}{\Xi\sqrt{\Delta_r}\sin\alpha}\left[\frac{\dot{\phi}\Sigma\Delta_r}{E\left(r^2+a^2\right)-a\Xi L}-a\Xi\right]\right|_{(r_0,\theta_0)},
\end{equation}
where $\dot{r}=d r/d\sigma$ and $\dot{\phi}=d\phi/d\sigma$. From Eqs. (\ref{dotphi}) and (\ref{dotr}), and by substituting the impact parameters (\ref{xi}) and (\ref{eta}), we finally obtain
\begin{equation}
\sin\alpha=\left.\frac{\sqrt{\left[\left(\xi(r_p)-a\right)^2+\eta(r_p)\right]\Delta_r}}{r^2+a^2-a\Xi \xi(r_p)}\right|_{(r_0,\theta_0)}
\end{equation}
and
\begin{equation}
\sin\beta=\left.\frac{\sqrt{\Delta_r}\sin\theta}{\sqrt{\Delta_\theta}\sin\alpha}\left[\frac{a-\Xi\xi(r_p)\csc^2\theta}{a\Xi\xi(r_p)-\left(r^2+a^2\right)}\right]\right|_{(r_0,\theta_0)}.
\end{equation}

In order to plot the boundary curve of the shadow, we consider the stereographic projection from the celestial sphere onto a plane, where the Cartesian coordinates are written in the form
\begin{equation}
x(r_p)=-2\tan\left(\frac{\alpha(r_p)}{2}\right)\sin\left(\beta(r_p)\right),
\end{equation}
\begin{equation}
y(x_p)=-2\tan\left(\frac{\alpha(r_p)}{2}\right)\cos\left(\beta(r_p)\right).
\end{equation}
In Fig. \ref{shadow1}, the shadow of the rotating black hole with cosmological constant $\Lambda_{4D}=10^{-7}$ is presented, as seen by an observer placed at $r_0=50$ and $\theta_0=\pi/2$, for different values of the rotation parameter $a$ and the tidal charge $q$. For a fixed value of $a$, the radius of the shadow decreases and the deformation of the silhouette grows with $q$; the smallest and most deformed case corresponds to the extremal value $q_c$. Negative values of $q$ enlarge the shadow with respect to the Kerr--(anti--)de Sitter black hole case (which is recovered if $q=0$), while positive ones diminish it. For a fixed value of the rotation parameter and the charge, the apparent shape changes with the value of the cosmological constant, as is shown in Fig. \ref{shadow2}, in which $\Lambda_{4D}=-10^{-4}$ (dotted line), $\Lambda_{4D}=10^{-7}$ (dashed line), and $\Lambda_{4D}=10^{-4}$ (full line). In this case, the size of the shadow increases as $\Lambda_{4D}$ gets smaller, resulting in larger shadows for negative $\Lambda_{4D}$.

\section{Observables}\label{obs}

\begin{figure}[t!]
\begin{center}
    \includegraphics[scale=0.9]{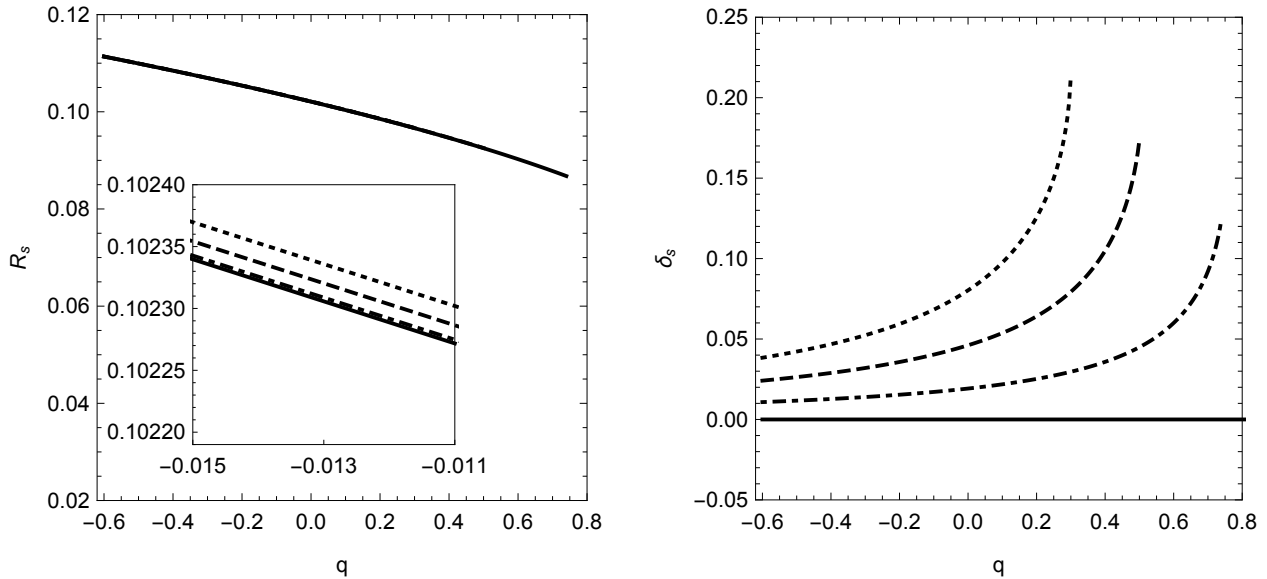}
    \vspace{0.25cm}
    \caption{Observables $R_s$ (left) and $\delta_s$ (right) as functions of $q$ for $\Lambda_{4D}=10^{-7}$: $a=0$ (full line), $a=0.5$ (dashed-dotted line), $a=0.7$ (dashed line), and $a=0.83$ (dotted line). In all cases, the observer is placed at $r_0=50$ and $\theta_0=\pi/2$.}
    \label{obs1}
\end{center}
\end{figure}

The co--rotating photons can get closer to the black hole, so the resulting shadow is displaced and compressed on this side, while the other side corresponding to counter--rotating photons is elongated. In order to characterize the apparent shape of the spinning black hole with a cosmological constant, we define two observables $R_s$ and $\delta _{s}$, analogously as it was done in Ref. \cite{hioki}. The parameter $R_s$ gives the approximate size of the shadow and is determined by the radius of a reference circle passing by three points: the top position $(x_t, y_t)$, the bottom position $(x _b,y _b)$, and the point corresponding to the unstable retrograde circular orbit as seen by an observer on the equatorial plane $(x _r,0)$. The distortion parameter $\delta _{s}$ measures the deformation of the shadow with respect to the reference circle; it is defined by $\delta _{s}=D/R_s$, with $D$ the difference between the endpoints of the circle and of the shadow, both of them at the opposite side of the point $(x _r,0)$, i.e. corresponding to the prograde circular orbit. In the case of the black hole in the Randall--Sundrum brane, for a given inclination angle $\theta _0$, the observables $R_s$ and $\delta_s$ are determined by four quantities: $\Lambda_{4D}$, $M$, $a$, and $q$. The gravitational effects on the shadow, which grow with $\theta _0$, are larger when the observer is situated in the equatorial plane of the black hole, i.e. when  $\theta_{0}=\pi/2$. In this case, the observables take the form
\begin{equation}
R_s=\frac{(x_t-x_r)^2+y^{2}_{t}}{2|x_t-x_r|}
\end{equation}
and
\begin{equation}
\delta_s=\frac{|\tilde{x}_p-x_p|}{R_s},
\end{equation}
where $(\tilde{x}_p, 0)$ and $(x_p, 0)$ are the points where the reference circle and the contour of the shadow cut the horizontal axis at the opposite side of $(x_r, 0)$, respectively. We have replaced the celestial coordinates adopted in Ref. \cite{hioki} by the Cartesian coordinates introduced in the previous section.

\begin{figure}[t!]
\begin{center}
    \includegraphics[scale=0.9]{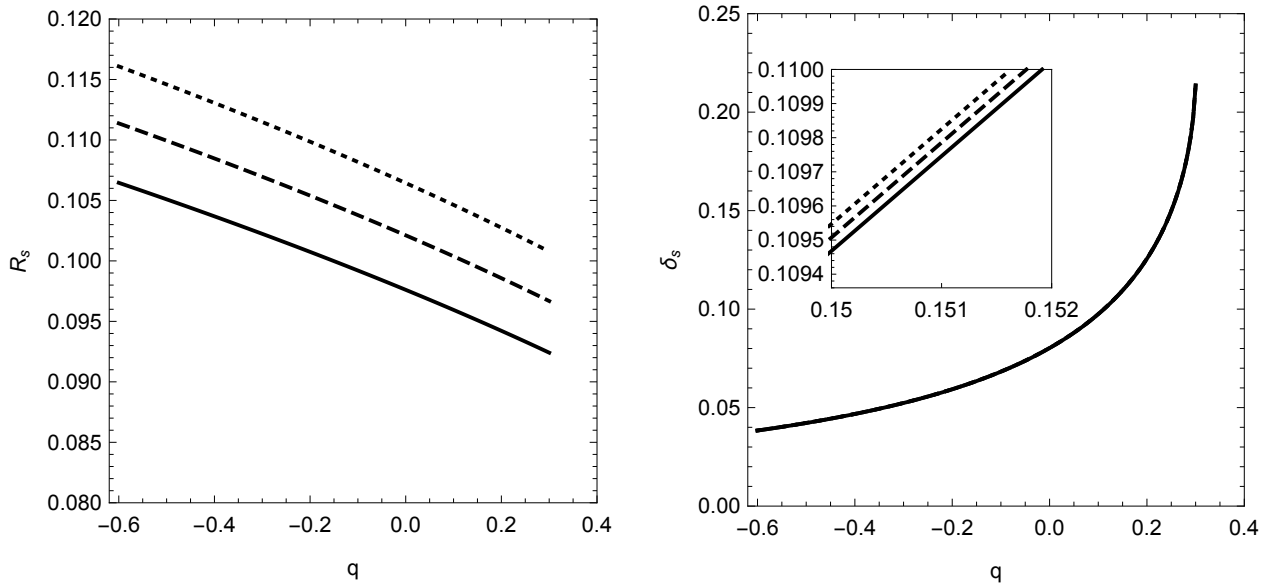}
    \vspace{0.25cm}
    \caption{Observables $R_s$ (left) and $\delta_s$ (right) as functions of $q$ for $a=0.83$, considering the observer placed at $r_0=50$ and $\theta_0=\pi/2$: $\Lambda_{4D}=-10^{-4}$ (dotted line), $\Lambda_{4D}=10^{-7}$ (dashed line), and $\Lambda_{4D}=10^{-4}$ (full line).}
    \label{obs2}
\end{center}
\end{figure}

As an example, we obtain the observables for an observer at $r_0=50$ with inclination angle $ \theta_0=\pi/2$.  In Fig. \ref{obs1}, the observables are shown as a function of $q$, for a fixed value of the cosmological constant $\Lambda_{4D}=10^{-7}$ and different values of the rotation parameter: $a=0$ (full line), $a=0.5$ (dashed-dotted line), $a=0.7$ (dashed line), and $a=0.83$ (dotted line). We see a weak dependence of the radius size $R_s$ with $a$, being a decreasing function of the charge, as can be seen in the frame inside, where a subrange of $q$ was taken. The distortion parameter $\delta_s$ is an increasing function of the charge and gets maximal when the charge reaches the limiting value $q_c$. For a fixed $q$, the observable $\delta_s$ increases with $a$. There is no distortion of the shadow for the static case, as expected. Negative values of $q$ lead to larger shadows (and then $R_s$) and smaller distortions with respect to the Kerr--(anti--)de Sitter black hole (corresponding to $q=0$). In Fig. \ref{obs2}, $R_s$ and $\delta_s$ are presented as functions of $q$ for $a=0.83$ for three values of the cosmological constant:  $\Lambda_{4D}=-10^{-4}$ (dotted line), $\Lambda_{4D}=10^{-7}$ (dashed line), and $\Lambda_{4D}=10^{-4}$ (full line). In this case, the radius of the apparent shape diminishes as $\Lambda_{4D}$ gets larger. The variation in the distortion $\delta_s$ are not noticeable for a fixed $q$, being slightly larger in the case of the smaller value of $\Lambda_{4D}$, with differences of order $10^{-5}$, as shown in the frame inside the figure.

\section{Final remarks}\label{remarks}

We have investigated the shadow of a rotating black hole in the Randall--Sundrum type II brane, with the presence of an effective four--dimensional cosmological constant \cite{neves}, so the geometry is not asymptotically flat. The rotating solution on the brane without cosmological constant was found a few years before \cite{aliev} and the corresponding shadow was also studied \cite{amarilla}. These geometries are characterized by the presence of a tidal charge, resulting from the influence of the bulk on the brane, which can be positive or negative, with the negative sign being more natural \cite{dadhich}. Here we have considered both de Sitter and anti--de Sitter asymptotics, and we have analyzed the effects produced by the cosmological constant and the tidal charge. Because of the presence of the cosmological constant, the observer cannot be at infinity and the apparent shape depend on the position of the observer, which is always in a non--flat region of spacetime. As a consequence, we have adopted the recently introduced approach for the shadow of the Kerr--Newman--NUT--(anti--)de Sitter black hole \cite{perlick}. We have presented two observables, one related to the apparent size and the other to the deformation of the shadow. For a given fixed position of the observer, by analyzing these observables, we have seen that larger values of the cosmological constant decrease the radius of the reference circle and diminish the distortion of the shadow. When fixing the value for the cosmological constant, the radius is a decreasing function of the tidal charge and increases with the rotation parameter. In contrast, the distortion increases with both the tidal charge or the rotation parameter.  A negative tidal charge enlarges the shadow of the black hole, but reduces the distortion due to the rotation. In our model there is vacuum in vicinity of the black hole, but if it is surrounded by a plasma, photons undergo various effects, such as absorption, scattering, and refraction, which depend of the specific characteristics of the medium involved. The  influence of the presence of plasma in the shadow was recently analyzed for the Kerr spacetime \cite{shadowplas}, the size and the shape of the shadow in this case depend on the photon frequency. Then, if the black hole considered in our work is surrounded by a plasma, chromatic effects on the shadow will appear.

The size and shape of the shadow depends directly on the properties of the black hole geometry involved, so it will serve to test general relativity and alternative theories. Nowadays, much attention is paid to observing the closest supermassive black holes. The instrument GRAVITY will examine in the near-infrared band the vicinity of the supermassive black hole in the center of our galaxy (Sgr A*) \cite{gillessen17}, following with high precision the orbits of the stars close to it. Through this monitoring, the measurement of the black hole mass will be improved, and the determination of the spin and the quadrupole moment may be also possible. The observation of the shadow associated to Sgr A* is one of the main goals to be reached in the near future, because of its large size and proximity \cite{observ}. Among the different projects for this purpose, one is the Event Horizon Telescope, which uses very long baseline interferometry to combine existing radio facilities into a telescope with very high angular resolution (of about 15 $\mu$as). The Event Horizon Telescope has already made observations of the center of the Milky Way and also of the nearby giant elliptic galaxy M87 \cite{broderick15}, and the first image of the shadow of a black hole is expected in short. Millimetron is a planned space-based mission operating from far infrared to millimeter wavelengths, with expected resolution of 0.05 $\mu$as. Forthcoming  x--ray instruments will also have an improved resolution that will allow a detailed exploration of the Galactic center. For more details about the observational prospects, see Ref. \cite{observ} and the references therein. The comparison between the observed apparent shape of black holes and the different theoretical models will be a useful tool in future astrophysics, but the subtle effects analyzed in our work seems to require a more advanced generation of instruments.

\section*{Acknowledgments}

This work has been supported by CONICET and  University of Buenos Aires.

\end{document}